

\magnification=1200
\hsize 6.0truein
\vsize 8.5truein
\parindent 1.0truecm
\font\smallrm=cmr8 scaled\magstephalf	

\rm
\null


\footline={\hfil}
\vglue 0.8truecm

\rightline{\bf DFUPG-67-93}
\rightline{\sl December 1993}
\vglue 2.0truecm

\centerline{\bf QCD, the Parton Model, and the Nucleon }
\centerline {\bf Polarised Structure Functions }
\vglue 2.0truecm

\centerline{ Paolo M. Gensini}
\centerline{\sl Dip. di Fisica dell'Universit\`a di Perugia, Perugia,
Italy, and }
\centerline{\sl Sezione di Perugia dell'I.N.F.N., Perugia, Italy}

\vglue 5.0truecm

\centerline{ Talk presented at }
\centerline{\sl Quinto Convegno su Problemi di Fisica Nucleare Teorica, }
\centerline{\sl Cortona (Arezzo), October 1993 }
\vglue 1.0truecm
\centerline{ To be published in the proceedings: }
\centerline{\sl ``Perspectives on Theoretical Nuclear Physics (V)'', }
\centerline{ L. Bracci, {\sl et al.}, eds. (ETS Ed., Pisa 1994) }
\pageno=0
\vfill
\eject

\footline={\hss\tenrm\folio\hss}	
\centerline{\bf QCD, the Parton Model, and the Nucleon }
\centerline {\bf Polarised Structure Functions }
\vglue 1.0truecm

\centerline{ Paolo M. Gensini}
\centerline{\sl Dip. di Fisica dell'Universit\`a di Perugia, Perugia,
Italy, and }
\centerline{\sl Sezione di Perugia dell'I.N.F.N., Perugia, Italy}

\vglue 1.0truecm
\centerline{\bf ABSTRACT }
\vglue 0.5truecm
{\narrower\smallskip\smallrm
\par The present talk summarises the 1993 situation in understanding
the spin structure of the nucleon via electron and muon polarised
deep--inelastic scattering (PDIS). The central question I shall address here
is if the data can be interpreted as evidence for polarisation in the
``strange'' nucleon ``sea'', and I conclude that they can not: incidentally,
I also find that they can not be constructed as evidence for violation of
perturbative QCD (PQCD), either.\smallskip}

\vglue 1.0truecm
\leftline{\bf 1. Introduction. }
\vglue 0.5truecm
\par In April 1975, the cover of the CERN Courrier represented the clamour
raised by the then recent discoveries of the $J/\psi$ and $\psi'$ by two,
radiative--tail--shaped piles of papers, the first and taller of which labeled
``Theory'', the second, much lower and thinner, ``Experiments'': such an
image can very well represent again today what has happened to our
community just after the 1988 publication by the EMC$^1$ of their
results on the
asymmetry in muon PDIS on a polarized hydrogen (but actually spin--frozen
ammonia) target.
\par As then, these last few years have also seen subsequent experiments and
deeper theoretical studies weeding out much of the ``Theory'' pile: in this,
forcefully short, review, I shall concentrate only on the most significant,
recent steps in this direction, and on their impact on data interpretation.
\par The rest of this paper will be divided in three parts: a presentation of
the most advanced evaluations of PQCD corrections to the first--moment sum
rules, a careful discussion of the original EMC experiment and of the problems
posed in general when going from the asymmetry $A$ to the polarized structure
function $g_1$, and then a discussion on the spin composition of the nucleon
and in particular on its ``strange'' component.
\vfill
\eject

\leftline{\bf 2. PQCD developments. }
\vglue 0.5truecm
\leftline{\sl 2.1. The Bjorken sum rule. }
\vglue 0.5truecm
\par The first--moment sum rule for the difference between proton and neutron
polarized structure functions reads$^2$ (written in its full QCD garb)
$$
\int_0^1 dx [ g_1^p(x,Q^2) - g_1^n(x,Q^2) ] = C(\alpha_s) \cdot {1\over6}
g_A + h.t.\ ,
\eqno (1)
$$
where $C(\alpha_s)$ are the PQCD corrections to the parton--model result,
and ``h.t.'' stands for higher--twist contributions$^3$ (HTC), behaving as
integer inverse powers of $Q^2$ for $Q^2 \to \infty$. Since recent experiments
extend from $<Q^2>\ = $ 2.0 GeV$^2$ (SLAC E142$^4$) to the 10.7 GeV$^2$ of
EMC$^{1,5}$, inclusion of these effects might be crucial to a test of
the Bjorken sum rule (BjSR), Eq. 1. This is indeed the point taken by Ellis
and Karliner$^6$ in their analisys of PDIS data.
\par An estimate of HTC has been given$^3$ -- and recently revised$^7$ -- by
Balitski\u\i, Braun and Kolesnichenko: two remarks are in order here.
First, any evaluation of HTC to a sum rule such as Eq. 1 must depend on the
order up to which $C(\alpha_s)$ is evaluated; second, the extrapolation away
from $Q^2 = 0$ of the Drell--Hearn--Gerasimov sum rule$^8$ suggests that HTC
``die off'' already at moderate values of $Q^2$ of the order of 1 GeV$^2$,
much sooner thus than in the corresponding, ``unpolarised'' case of the
Gottfried sum rule. This latter fact is substantiated by strong cancellation
present -- although between different--order terms -- in the estimate of refs.
(3,7).
\par I shall discuss in the rest of this sub--section the recent
improvement$^9$ in the evaluation of $C(\alpha_s)$ and its implications. This
coefficient is now known to $O(\alpha_s^3)$ and can be written as
$$
C(\alpha_s) = 1 - {\alpha_s\over\pi} - c_1(N_f) ({\alpha_s\over\pi})^2 -
c_2(N_f) ({\alpha_s\over\pi})^3 - ...\ ,
\eqno (2)
$$
where the values of $c_{1,2}$ for $N_f$ from 3 to 5 are listed in Table I,
copied from the original paper by Larin and Vermaseren$^9$. Due to the large
values of the coefficients $c_{1,2}$, $C(\alpha_s)$ is evidently converging
very slowly at low values of $Q^2$, such as those of SLAC experiment E142$^4$
and of the SMC$^{10}$ ($<Q^2>\ =$ 4.6 GeV$^2$).
\par A technique for an estimate of the magnitude of higher, not yet computed
terms is the so--called ``accelerated convergence'', consisting in turning the
power series into a continued fraction (or Pad\'e approximant), for which
convergence is rigorous only if the function expanded were a Stieltjes one
(which is probably {\sl not} the case for PQCD, but a test on the 2nd--order
approximation to Eq. 2 shows that, despite this fact, the technique
works reasonably well),
namely in replacing Eq. 2 with
$$
C(\alpha_s)^{-1} = 1 + {\alpha_s\over\pi} [ 1 - [ 1 + c_1(N_f)]
{\alpha_s\over\pi}
[ 1 - {{c_2(N_f) - c_1(N_f)^2}\over{1 + c_1(N_f)}} {\alpha_s\over\pi} [ 1 -
... ]^{-1} ]^{-1} ]^{-1} \ ,
\eqno (2')
$$
which one can expect to approximate the ``full'' $C(\alpha_s)$ better then Eq.
2 at the lowest values of $Q^2$. Accordingly, one {\sl must} use here the
3--loop expansion for $\alpha_s(Q^2)$ and its scale
$\Lambda_{\overline{MS}}^{(N_f)}$:
a recent PQCD analysis by Bethke and Catani$^{11}$ of {\sl all} available data
(both space--like and time--like) leads to a {\sl conservative} estimate of
$\Lambda_{\overline{MS}}^{(5)}$ = 200 $\pm$ 50 MeV, and therefore to the value
$\Lambda_{\overline{MS}}^{(3)}$ = 411 $\pm$ 103 MeV that will be used
throughout.

\vglue 0.6truecm
\centerline{\bf Table I}
\centerline{\bf Coefficients of higher QCD corrections to BjSR }
\vglue 0.3truecm
\hrule
$$\vbox{\halign{#\hfil&\qquad\hfil#\hfil&\qquad\hfil#\hfil\cr
$N_f$ & $c_1(N_f)$ & $c_2(N_f)$ \cr
3 & 3.5833 & 20.2153 \cr
4 & 3.2500 & 13.8503 \cr
5 & 2.9167 & 7.8402 \cr}}$$
\hrule
\vglue 0.6truecm

\par A question better addressed at this point is the actual value of $N_f$,
the number of ``active'' flavours, to be used at each $<Q^2>$: in
annihilation,
and in general for time--like $Q^2$, there is no ambiguity, since flavour
thresholds can be reasonably set for heavy quarks at $Q^2\simeq4m_Q^2$. In
deep--inelastic scattering (DIS) a flavour is ``active'' {\sl only} when {\sl
appreciably} contributing to the moment sum rules, i.e. when produced a) in a
{\sl really inclusive} manner (read: not only in low--multiplicity events),
and b) over an appreciable range in Bjorken's variable $x$ (say up to $x
\simeq 1/3$). If one sets the beginning of the scaling region at $Q^2 \simeq$
2 GeV$^2$ (as indicated by the ``classic'' SLAC--MIT experiments), the
previous requirements ask for a $Q^2 \ge$ 17 GeV$^2$ for charm to be an
``active'' flavour in DIS. The choice of $N_f = 3$ (rather than 4) has no
great effect on $C(\alpha_s)$, as one can read from Table I, but for
the $Q^2$--evolution of the unitary--singlet piece the coefficient of
the 1st--order term in Eq. 2 almost cancels, for $N_f = 4$, the
1st--order one, $N_f/\beta_0$, coming from the
anomalous dimension of the anomaly$^{12}$. The
1st--order formul\ae\ used by Preparata and Ratcliffe$^{13}$ are thus plainly
wrong (not even mentioning their peculiar -- to say the least --
interpretation of the anomaly).
\par A quick summary of this sub--section is given in Table II, where we list
the
PQCD coefficients truncated at the $m$--th power for $m =$ 1 and 3,
$C(\alpha_s)_m$, together with the accelerated--convergence estimate
$C(\alpha_s)_{ac}$, all evaluated with the 3--loop expression for
$\alpha_s(Q^2)$ as in ref. 11: the reader can gauge by her/himself
the impact of higher terms at the lowest $Q^2$--values.
\par A comparison with the BjSR integral, as evaluated by ref. 6 and by Close
and Roberts$^{14}$, shows that HTC can indeed be only a minor correction at
$Q^2 \ge $ 2 GeV$^2$, provided one treats with adequate care the PQCD
correction factor $C(\alpha_s)$ -- or, if one likes it better, HTC
can be easily
traded for a slight modification of the latter, such as provided by an
``accelerated convergence'' expansion.

\vglue 0.6truecm
\centerline{\bf Table II}
\centerline{\bf The PQCD factor of BjSR }
\vglue 0.3truecm
\hrule
$$\vbox{\halign{#\hfil&\qquad\hfil#\hfil&\qquad\hfil#\hfil&\qquad\hfil#\hfil&
\qquad\hfil#\hfil\cr
$Q^2$ & $\alpha_s$ & $C(\alpha_s)_1$ & $C(\alpha_s)_3$ &
$C(\alpha_s)_{ac}$ \cr
12.0 & 0.2395 & 0.9238 & 0.8940 & 0.8876 \cr
10.7 & 0.2454 & 0.9219 & 0.8904 & 0.8832 \cr
 8.0 & 0.2619 & 0.9166 & 0.8800 & 0.8702 \cr
 6.0 & 0.2809 & 0.9106 & 0.8675 & 0.8535 \cr
 4.6 & 0.3018 & 0.9039 & 0.8529 & 0.8330 \cr
 3.0 & 0.3445 & 0.8903 & 0.8206 & 0.7804 \cr
 2.0 & 0.4015 & 0.8722 & 0.7714 & 0.6726 \cr}}$$
\hrule

\vglue 1.0truecm
\leftline{\sl 2.2. The isosinglet sum rule and the anomaly's anomalous
dimension. }
\vglue 0.5truecm
\par In 1974, Ellis and Jaffe$^{15}$, faced with the problem of how to use a
sum rule akin to Eq. 1 with only hydrogen data (and polarized targets
different from H$_2$ were over 18 years in the future), used parton--model
ideas to derive a sum rule for the 1st moment of $g_1^p$ alone; the
QCD--corrected version of such a sum
rule is a part of PQCD as fundamental as BjSR, and is best written for the
isoscalar combination od PDIS structure functions as
$$
\int_0^1 dx [ g_1^p(x,Q^2) + g_1^n(x,Q^2) ] = C(\alpha_s) \cdot [ {1\over{18}}
g_8 + {2\over9} g_0(Q^2) ] + h.t. \ .
\eqno (3)
$$
\par Additional complications with respect to the isovector one, Eq. 1, arise
from the facts a) that the isoscalar axial charges of the
nucleon are not directly measurable, and at least $g_8$ is indeed better known
via flavour--symmetry arguments ({\sl modulo}
symmetry--breaking effects) than through experiment (elastic neutrino--nucleon
or parity--violating electron--nucleon scattering), and b) that the
unitary--singlet one $g_0$ couples to the gluonic fields via the axial anomaly
and possesses therefore anomalous dimensions$^{12}$, so that its evolution
with $Q^2$ is not exausted by the PQCD factor $C(\alpha_s)$ and must be
explicitly computed.
\par Since the coupling is scheme--dependent, this point has been the centre
of a very heated theoretical debate. To cut a long history short, one can
summarise it by saying that, in the conventional parton language where the
masses of the partons are neglected with respect to the momentum scale $Q$,
one can put$^{16}$
$$
g_0(Q^2) = \sum_{i=1}^{N_f} \Delta q_i - N_f {\alpha_s\over{2\pi}} \Delta
G(Q^2) \ ,
\eqno (4)
$$
where $\Delta G$ is the first moment of the gluon polrised distribution
function $\delta G(x) = G_+(x) - G_-(x)$, and determine its evolution
via the equation (where $t = {\rm log} Q^2/\mu^2$)
$$
{d\over{dt}} g_0(t) = - N_f {\alpha_s\over{2\pi}} \gamma_{gq}(\alpha_s) g_0(t)
\ ,
\eqno (5)
$$
which relates to the anomalous dimension of the axial anomaly $\gamma_{gg}(
\alpha_s)$ via
$$
- N_f {\alpha_s\over{2\pi}} \gamma_{gq}(\alpha_s) = \gamma_{gg}(\alpha_s) -
\beta(\alpha_s) {{2\pi}\over\alpha_s}
\eqno (6)
$$
and gives, after integrating in $\alpha_s$,
$$
{\rm log} \ {{g_0(Q^2)}\over{g_0(\mu^2)}} = {{6N_f}\over{33-2N_f}}
{{\alpha_s(Q^2)-\alpha_s(\mu^2)}\over\pi} \ [ 1 +
$$
$$
+ \ ({{83}\over{24}} + {N_f\over{36}} -
{{33-2N_f}\over{8(153-19N_f)}}) \ {{\alpha_s(Q^2)+\alpha_s(\mu^2)}\over\pi}
+ ... ] \ ,
\eqno (7)
$$
with the 3--loop calculation by Larin$^{17}$ of the anomalous dimension.
\par As one can read from the above expression, $g_0(Q^2)$ can be drastically
reduced (still at $N_f = 3$) from its value at the scale $\mu^2$ for $Q^2 >
\mu^2$; on the other hand, one can not set $\mu\to\infty$ and thus drop
$\alpha_s(\mu^2)$ from Eq. 7, since the anomaly contribution to Eq. 4 is not
definable in this limit$^{16}$ due to $\Delta G(Q^2)$ growing asymptoyically
as $\alpha_s(Q^2)^{-1}$. It can also be noted that $g_0 = 0$ is a special
value, being a {\sl fixed point} in the evolution equation$^{16}$,
but on the other hand this does not allow to infer, for the same reason
that gives the evolution in Eq. 7, that $\sum_i \Delta q_i = 0$!
\par Sometimes, it has also been stated in the literature that the anomaly
contributes to $g_1(x)$ only at very small $x$ values: this is true (in the
above, $m_i = 0$ scheme) {\sl only if} the polarised gluon distribution {\sl
peaks} at $x = 0$, as e.g. in the ``intrinsic'' gluon distribution derived by
Brodsky and Schmidt$^{18}$. Unfortunately, the distribution they propose for
$\delta G(x)$ has the {\sl wrong} Regge behaviour for $x \to
0$ ({\sl contrary to their statement}), since it would require
dominance of $\delta G$ by the {\sl pion trajectory} with intercept
$\alpha_\pi(0)\simeq0$; instead, one would rather expect it to be dominated
by the pseudoscalar--glueball trajectory with intercept $\alpha_G(0)\le-1$.
When this constraint is imposed on the Brodsky--Schmidt framework,
{\sl ceteris paribus}, more than 80\% of
the anomaly contribution falls in the $x$ interval covered e.g. by the EMC
experiment, making the anomaly contribution virtually indistinguishable from
the other, ``intrinsic'', ``sea'' distributions, while the integral $\Delta G$
remains of order unity, as in ref. 17, or less, being rather solidly
tied to the
momentum fraction $<x_G>\ \simeq1/2$ carried by the gluons at the momentum
scale
at which these ``intrinsic'' components are defined.
\vfill
\eject
\leftline{\bf 3. Measurements and parametrizations for $g_1(x)$. }
\vglue 0.5truecm
\par What is actually measured are not the PDIS structure functions
themselves, but
rather the polarisation asymmetries $A = (\sigma^{\uparrow\uparrow} -
\sigma^{\uparrow\downarrow})/(\sigma^{\uparrow\uparrow} +
\sigma^{\uparrow\downarrow})$, related to the former by
$$
g_1(x,Q^2) = {{A \cdot F_2(x,Q^2)}\over{2x\cdot [1 + R(x,Q^2)]}} \ .
\eqno (8)
$$
\par It is obvious that the factor $2x$ in the denominator makes the direct
determination of $g_1(x)$ at $x\to0$ impossible for finite--accuracy data. The
evaluation of the sum rules, Eqs. 1 and 3, depends thus on the parametrisation
assumed to extrapolate $g_1(x)$ to $x=0$: the usual treatments of this point
have till now assumed it to extrapolate {\sl smoothly} to a constant as
$g_1(x)\simeq\alpha+\beta x$, in accord with the pion--pole trajectory
intercept
being close to zero. However, since one does not expect the ``sea''
distributions to couple dominantly to an isovector, pseudoscalar trajectory,
such as the pion,
but rather to an isoscalar one such as the eta, I shall rather have for
them
a behaviour $x^{-\alpha_\eta(0)}$, with $\alpha_\eta(0)\simeq1/4$, which,
together with the negative sign of the ``sea'' contribution, produces a spike
{\sl in the isoscalar part} of $g_1(x)$ at $x = 0$ of the type
$g_1(x)\simeq\alpha-\beta x^{-1/4}$ (with $\alpha$, $\beta>0$),
{\sl presently} unseen in the asymmetry $A$
just because of the $2x$ factor and the {\sl very limited} accuracy of the
data. Of course, such a ``spiky'' behaviour does not show in the integrand of
the BjSR, which can therefore be extrapolated smoothly to $x=0$ according to
the conventional practice in this matter.
\par Apart from this point, which has however a non--negligible influence on
the evaluation of the isoscalar sum rule, Eq. 3, raising the low--$x$
contribution to its l.h.s. well over the ``conventional'' estimates, one has
to avoid using the EMC published data$^5$ for $g_1^p$, and to use instead only
their values for $A$, with an {\sl adequate} set of values for $F_2$ and $R$.
\par Indeed, the EMC calculated originally$^1$ $g_1^p$ using a) the ratio
$R$ predicted by PQCD, systematically smaller than experiment since
finite--mass corrections are dominant at low values of $Q^2$ (the effect of
this is however not too important at $Q^2=$ 10.7 GeV$^2$), and b) {\sl their}
values for the {\sl unpolarized} structure function $F_2^p$, systematically
lower than those by the BCDMS collaboration$^{19}$ (and than the recent NMC
data$^{20}$ as well) by as much as 13\% at the lowest values of $x$.
\par Even using their {\sl measured} values of $A$ together with a
phenomenological parametrization for $F_2^p(x,Q^2)$ (and $R^p(x,Q^2)$) to
produce $g_1^p(x,Q^2)$ at a reference, fixed value of $Q^2$ (and assuming $A$
to vary little$^{6,14}$ with $Q^2$ --a yet to be proven assumption, in the
light of the sparse nature of available data) is not free of the above, last
source of error: indeed only the latest, post--NMC parametrizations$^{21}$
have dropped the {\sl unpolarized} EMC data altogether, while {\sl all}
previous analyses ended up averaging over the two, conflicting sets of
data for small values of $x$, as did the EMC in the full--paper version of
their work$^5$.
\vfill
\eject
\leftline{\bf 4. The isosinglet sum rule and the spin content of the nucleon. }
\vglue 0.5truecm
\par In the following table we present a re--evaluation of the integrals over
the three experiments$^{4,5,10}$ on the PDIS structure functions $g_1$, with
a {\sl personal} renormalization of the EMC $g_1^p(x)$ data, following the
prescriptions outlined in the previous section. There is an evident increase
in the proton integral over the EMC evaluation of ref. 5: the same behaviour
at $x = 0$ has been assumed for all three PDIS structure functions, and the
same increase is present for them as well for the low--$x$ part of the
integrals. Due to the
different nature of the targets and momentum scales we correct each integral
with the appropriate BjSR contribution, where we use the $C(\alpha_s)$
given by ``accelerated convergence'' (Eq. 2), and the value $g_A$ = 1.2555
$\pm$ 0.0015 from an overall analysis of {\sl all} baryon semileptonic
decay data$^{22}$.

\vglue 0.6truecm
\centerline{\bf Table III }
\centerline{\bf The isosinglet sum rule evaluation and analysis }
\vglue 0.3truecm
\hrule
$$\vbox{\halign{#\hfil&\quad\hfil#\hfil&\quad\hfil#\hfil&\quad\hfil#\hfil\cr
Experiment & revised EMC$^5$ & SMC$^{10}$ & SLAC E142$^4$ \cr
data $x$--range & 0.01 -- 0.70 & 0.04 -- 0.070${^\dagger}$ & 0.03 -- 0.70 \cr
from data & 0.129 $\pm$ 0.013 & 0.039 $\pm$ 0.023 & -0.019 $\pm$ 0.008 \cr
from low--$x$ & 0.006 $\pm$ 0.001 & 0.005 $\pm$ 0.003 & -0.005 $\pm$ 0.002 \cr
from high--$x$ & 0.001 & 0.001 & 0.000 \cr
$\int_0^1dxg_1$ & 0.136 $\pm$ 0.013 & 0.045 $\pm$ 0.023 & -0.024 $\pm$ 0.008
\cr
BjSR & 0.1848 $\pm$ 0.0002 & 0.1743 $\pm$ 0.0002 & 0.1407 $\pm$ 0.0002 \cr
isoscalar SR & 0.087 $\pm$ 0.026 & 0.090 $\pm$ 0.046 & 0.093 $\pm$ 0.016 \cr
$g_0(\alpha_s)/g_0(1)$ & 0.6820 & 0.6953 & 0.7220 \cr
$g_0(1)$ for $g_8$ = 0.75 & 0.376 $\pm$ 0.194 & 0.430 $\pm$ 0.358 & 0.602
$\pm$ 0.149 \cr
$g_0(1)$ for $g_8$ = 0.60 & 0.431 $\pm$ 0.194 & 0.484 $\pm$ 0.358 & 0.654
$\pm$ 0.149 \cr}}$$
\hrule
\noindent{\smallrm \dag) range reduced to avoid screening effects on the
very--low--$x$ data points. }
\vglue 0.6truecm

\par Here we derive $g_0(\alpha_s=1)$ either using the flavour $SU(3)$
prediction $g_8=3g_A^{\Xi\Lambda}\simeq$ 0.75 or the perhaps better value
$g_8\simeq$ 0.60, which includes a simple modeling of
$SU(3)$--symmetry--breaking effects in the baryon octet$^{23}$ (reproducing
the {\sl increase} in $\Sigma_{\pi N}$ over the quark--model prediction {\sl
without} a large ``strange sea'' contribution$^{24}$).
\par The reduction of $g_0(\alpha_s=1)$ with respect to $g_8$ can be explained
by the gluonic anomaly without any recourse to a ``strange'' component in the
spin density: for $g_8=$ 0.750 we expect indeed $g_0(\alpha_s=1)=g_8-{3\over
{4\pi}}(1-g_8)=$ 0.690 (in somewhat poor agreement with line 9 of Table III),
and, for the smaller value $g_8=$ 0.600,
$g_0(\alpha_s=1)=$ 0.505, in good accord with the results of the last line in
the table,
and leading to a rather small ``intrinsic'' value for $\Delta G(\alpha_s=1)$.
\vfill
\eject
\leftline{\bf 4. Summary and conclusions. }
\vglue 0.5truecm
\par The three experiments on the PDIS {\sl asymmetries} (SLAC experiments
E142$^4$, and the collaborations EMC$^{1,5}$ and SMC$^{10}$ at CERN)
do not contradict conventional expectations on the
spin structure of the nucleon, namely that one should have a ``strange'' spin
component very close to zero on one side,
but on the other no pure valence--quarks either, as
known since more than twenty years (and a complete list of references
would be as long as this paper itself:
I just refer to the recent, illuminating papers by Lipkin$^{25}$)
from the reduction in $g_A$ with respect to
the quark--model prediction $g_A = 5/3$.
\par One finds a drastic reduction in $g_0(Q^2)$ from the
parton--model--plus--OZI--rule expectation $g_0 \simeq g_8$, due {\sl both} to
the presence of the QCD axial anomaly {\sl and} to its anomalous dimension
(and, as one can read from Table III, due more to the second than to the first
reason): a {\sl correct} use of QCD with high orders included (which make
these effects even larger than 1st--order alone) is necessary to describe the
first--moment sum rules without conflict with experiments (and our
expectations). The ``spin crisis'' of 1988 was the result of inadequate
theoretical description, as much as of a somewhat low normalization in the
$F_2$ values used by the EMC.
\par Last but not least, HTC are not needed to explain the data, at least for
$Q^2 \ge$ 2 GeV$^2$ (the SLAC E142 average momentum scale), though essential
in connecting the BjSR, Eq. 1, to the Drell--Hearn--Gerasimov sum rule at
$Q^2 = 0$.

\vglue 1.0truecm
\centerline{\bf REFERENCES}
\vglue 0.5truecm

\item{1.} {The European Muon Collaboration (J. Ashman, {\sl et al.}): {\sl
Phys. Lett.} {\bf B 206} (1988) 364.}
\item{2.} {J.D. Bjorken: {\sl Phys. Rev.} {\bf 148} (1966) 1467; {\sl Phys.
Rev.} {\bf D 1} (1970) 1376.}
\item{3.} {Ya.Ya. Balitski\u\i, V.M. Braun and A.V. Kolesnichenko: {\sl JETP
Lett.} {\bf 50} (1989) 61; {\sl Phys. Lett.} {\bf B 242} (1990) 245.}
\item{4.} {The E142 Collaboration (P.L. Anthony, {\sl et al.}): {\sl Phys.
Rev. Lett.} {\bf 71} (1993) 959.}
\item{5.} {The European Muon Collaboration (J. Ashman, {\sl et al.}): {\sl
Nucl. Phys.} {\bf B 328} (1989) 1.}
\item{6.} {J. Ellis and M. Karliner: {\sl Phys. Lett.} {\bf B 313} (1993) 131;
see also their rapporteur's talk (pres. by J. Ellis) to {\sl ``PANIC
XIII''}, Perugia June--July 1993, {\sl report CERN--TH. 7022/93}
(Geneve 1993).}
\item{7.} {Ya.Ya. Balitski\u\i, V.M. Braun and A.V. Kolesnichenko: errata to
ref. 3, {\sl DESY report} (Hamburg 1993), subm. to {\sl Phys. Lett.} {\bf
B}.}
\item{8.} {R.L. Workman and R.A. Arndt: {\sl Phys. Rev.} {\bf D 45} (1992)
1789; V.D. Burkert and B.L. Ioffe: {\sl Phys. Lett.} {\bf B 296} (1992)
223; V.D. Burkert and Z.--J. Li: {\sl Phys. Rev.} {\bf D 47} (1993) 46;
V. Bernard, N. Kaiser and U.--G. Mei{\ss}ner: {\sl Phys. Rev.} {\bf D 48}
(1993) 3062.}
\item{9.} {S.A. Larin and J.A.M. Vermaseren: {\sl Phys. Lett.} {\bf B 259}
(1991) 345.}
\item{10.} {The Spin Muon Collaboration (B. Adeva, {\sl et al.}): {\sl Phys.
Lett.} {\bf B 302} (1993) 533.}
\item{11.} {S. Bethke and S. Catani: {\sl report CERN--TH. 6484/92} (Geneve
1992), summary pres. at the {\sl ``27th Rencontre de Moriond''}, Les Arcs
March 1992.}
\item{12.} {J. Kodaira: {\sl Nucl. Phys.} {\bf B 165} (1980) 129.}
\item{13.} {G. Preparata and P.G. Ratcliffe: {\sl ``EMC, E142, SMC, Bjorken,
Ellis--Jaffe ... and All That''}, {\sl Univ. di Milano report} (Milano
1993), and references therein.}
\item{14.} {F.E. Close and R.G. Roberts: {\sl Phys. Lett.} {\bf B 316} (1993)
165; see also the talk presented by F.E. Close at the {\sl ``6th ICTP
Workshop''}, Trieste May 1993, {\sl report RAL 93--034} (Chilton 1993).}
\item{15.} {J. Ellis and R.L. Jaffe: {\sl Phys. Rev.} {\bf D 9} (1974) 1444;
{\sl Phys. Rev.} {\bf D 10} (1974) 1669.}
\item{16.} {G. Altarelli and B. Lampe: {\sl Z. Phys.} {\bf C 47} (1990) 315.}
\item{17.} {S.A. Larin: {\sl Phys. Lett.} {\bf B 303} (1993) 113.}
\item{18.} {S.J. Brodsky and I. Schmidt: {\sl Phys. Lett.} {\bf B 234}
(1990) 144.}
\item{19.} {A.C. Benvenuti, {\sl et al.}: {\sl Phys. Lett.} {\bf B 223}
(1989) 485; {\sl Phys. Lett.} {\bf B 237} (1989) 592, 599.}
\item{20.} {The New Muon Collaboration (P. Amaudruz, {\sl et al}): {\sl Nucl.
Phys.} {\bf B 371} (1992) 3; {\sl Phys. Lett.} {\bf B 295} (1992) 159.}
\item{21.} {A.D. Martin, W.J. Stirling and R.G. Roberts: {\sl Phys. Rev.}
{\bf D 47} (1993) 867; {\sl Phys. Lett.} {\bf B 306} (1993) 145, {\sl
erratum} {\bf B 309} (1993) 492; H. Plothow--Besch: {\sl Comput. Phys.
Commun.} {\bf 75} (1993) 396.}
\item{22.} {P.M. Gensini: {\sl report DFPUG--66--93} (Perugia 1993), to be
publ. in $\pi N$ {\sl Newslett.}.}
\item{23.} {P.M. Gensini: {\sl Nuovo Cimento} {\bf A 103} (1990) 303.}
\item{24.} {P.M. Gensini: {\sl Nuovo Cimento} {\bf A 102} (1989) 75, {\sl
erratum} 1181; {\sl Nuovo Cimento} {\bf A 103} (1990) 1311; $\pi N$ {\sl
Newslett.} {\bf 6} (1992) 21.}
\item{24.} {H.J. Lipkin: {\sl Phys. Lett.} {\bf B 237} (1990) 130; {\sl Phys.
Lett.} {\bf B 251} (1990) 613.}

\bye